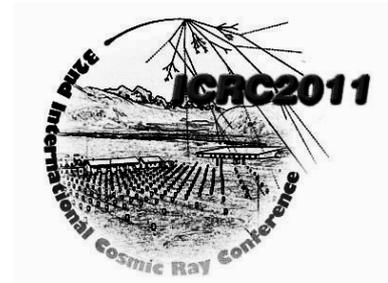

# Optical Performances of Slewing Mirror Telescope for UFFO-Pathfinder

S. JEONG[1], K. -B. AHN[2], J. W. NAM[1], I. H. PARK[1], S. -W. KIM[2], H. S. CHOI[3], Y. J. CHOI[4], B. GROSSAN[5], I. HERMANN[4], A. JUNG[1], Y. W. KIM[1], J. E. KIM[1], E. V. LINDER[1,5], J. LEE[1], H. LIM[1], K. W. MIN[4], G. W. NA[1], K. H. NAM[1], M. I. PANASYUK[6], G. F. SMOOT[1,5], S. SVERTILOV[6], Y. D. SUH[4], N. VEDENKIN[6], I. YASHIN[6], M. H. ZHAO[1], FOR THE UFFO COLLABORATION
[1]*Ewha Womans University, Seoul, Korea*
[2]*Yonsei University, Seoul, Korea*
[3]*Korea Institute of Industrial Technology, Ansan, Korea*
[4]*Korea Advanced Institute of Science and Technology, Daejeon, Korea*
[5]*University of California, Berkeley, USA*
[6]*Moscow State University, Moscow, Russia*
*soominjeong@gmail.com*

**Abstract:** The Ultra-Fast Flash Observatory-Pathfinder (UFFO-P) is to be launched onboard *Lomonosov* spacecraft in November 2011. It is to measure early UV/Optical photons from Gamma Ray Bursts (GRBs). Slewing Mirror Telescope (SMT) is one of two instruments designed for detection of UV/Optical images of the GRBs. SMT is a Ritchey-Chrétien telescope of 100 mm in diameter with a motorized slewing mirror at the entrance providing 17×17 arcmin$^2$ in Field of View (FOV) and 4 arcsec in pixel resolution. Its sky coverage can be further expanded up to 35 degrees in FOV by tilting a motorized slewing mirror. All mirrors were fabricated to about RMS 0.02 waves in wave front error (WFE) and 84.7% (in average reflectivity) over 200nm~650nm range. SMT was aligned to RMS 0.05 waves in WFE (test wavelength 632.8nm). From the static gravity test result, SMT optics system is expected to survive during launch. The technical details of SMT assembly and laboratory performance test results are reported

Keywords: **Gama Ray Bursts, UFFO, Slewing Mirror Telescope, Optical Performance.**

## 1  Introduction

Since the first discovery of GRB by Vela satellite in 1967 [1], several space missions have brought improved understanding to GRP physics. Especially, *Swift* observatory [2] launched in November 2004, has localized accurately the counterparts of GRBs in multi-wavelength regions. With such measurement, it has helped clearer understanding of optical emission from GRBs [3].

While the mission localized hundreds of GRBs longer than 60 seconds after the trigger and its instrument is faster than typical ground-based telescope, it has fundamental limitations to prompt GRB optical measurement earlier than 60 seconds. Like all other well-known GRB observing systems, it has to move the whole satellite and therefore is slow in targeting. This gives rise to urgent necessity of a new and faster instrument capable of GRB measurement earlier than 60 seconds after the trigger.

We have developed a new space telescope; the Ultra Fast Flash Observatory (UFFO) to explorer a sub-minute time scale of optical emission measurement from GRBs [4]. UFFO has a unique capability of rapid slewing with a scan mirror to redirect the incoming optical beam from targets. For the proof of rapid slewing mirrors concept to sub-minute measurement of GRB optical counterparts, a pathfinder project of UFFO (UFFO-P) has been developed as a part of the *Lomonosov* mission, scheduled to be launched in November 2011. Although UFFO-P mass and power are limited to less than 20kg and 20W, it will lay the foundation of technical feasibility for the main UFFO instrument.

UFFO-P consists of two instruments (Figure 1): the UFFO Burst Alert Telescope (UBAT) [5], and the Slewing Mirror Telescope (SMT) [6, 7]. The UBAT is a 1.85 sr FOV coded aperture mask gamma-camera using 1296 LYSO+MAPMTs. It provides SMT with trigger signals of the event and localization within 10 arcmin (7σ precision) in energy range of 5-200 k$e$V. The SMT is a 10cm aperture telescope with an Intensified Charge-Coupled Device (ICCD) detector for position refinement to 1 sec in UV/Optical region.

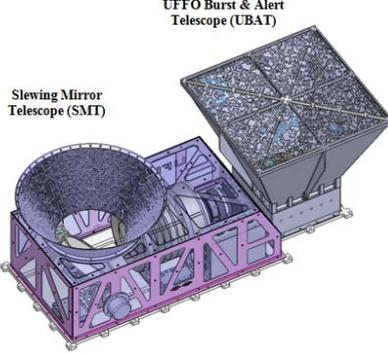

Figure 1. A rendering of integrated UFFO-P

This paper is concerned with the SMT optics system including design, construction, and laboratory test results in the following section 2. In section 3, we provide concluding remarks.

## 2  SMT Optical System

### 2.1 Optical Requirements

The SMT optical system requirements are listed in Table 1 below. The SMT optical system is designed for sufficient stiffness within the tight mass budget, while enjoying the thermal stability to deliver the required optical performances in operation.

| Characteristics | Requirement (Motorized slewing mirror / Telescope) |
|---|---|
| Mass (kg) | 1.8 / 1 |
| Dimension (mm) | 160×320×210 / 180×180×250 |
| Operation temp. (˚C) | -20 ~ 40 |
| Launch load (g) | 60 |
| Resonance freq.(Hz) | > 50 |
| Wavelength (nm) | 200 ~ 650x |
| Field of view (arcdeg) | 35×35 / 0.142×0.142 |
| Target pixel size (μm) | 22.2 |
| Nyquist sampling freq. (mm$^{-1}$) | 22.52 |
| Modulation Transfer Function (%) | > 0.5 |
| RMS WFE (wave) | 0.25 |

Table 1. SMT Optical Requirements

### 2.2 Motorized Slewing Mirror

UFFO-P has a motorized slewing mirror in front of a Ritchey Chrétien telescope to be pointed at the event within 1 sec. (Figure 2). Its tilt range can be expanded to ±17.5 degrees for 35 degrees in sky coverage (i.e. the half coded FOV of UBAT).

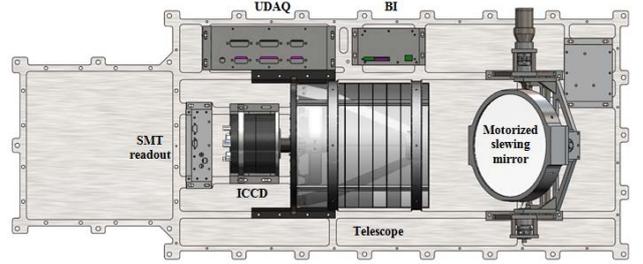

Figure 2. Overview of SMT instrument design

For the tight mass budget, the 6 inch slewing mirror is light weighted to 482g, achieving the lightweighting factor of about 57% while keeping the high stiffness. The slewing mirror is supported with pads made of RTV566 in its back, forward and side surfaces. The slewing mirror surface is coated with $SiO_2$, ensuring 13.6 nm RMS WFE and 84.7% average reflectivity in 200nm~700nm wavelength range.

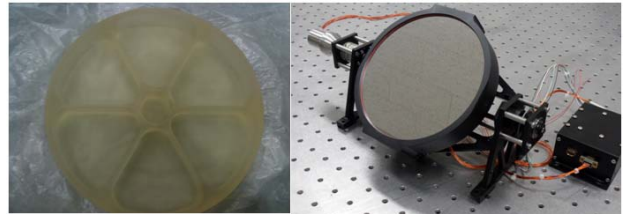

Figure 3. Backside of the 6 inch slewing mirror (left) and Manufactured motorized slewing mirror (right)

We use stepping motors and Harmonic Drive gears with 100:1 reduction ratio, providing the minimum step size of rotation to be 4.05 arcsec. A closed loop control with rotary encoders (GPI 7700) provides 0.7 arcmin of accuracy.

### 2.3 Telescope

● Optics Design

SMT is an aplanatic two mirror system with f/11.4 Ritchey-Chrétien type for compact design. It provides coverage of 17×17 arcmin$^2$ in field of view to the microchannel plate (MCP) Intensified Charge-Coupled Device (ICCD) detector. The MIC (MCP+ICCD) detectors cotain 256×256 pixels and each pixel corresponds to 4×4 arcsec² on the sky. To localize the optical counter part of GRBs effectively, a diffraction limited optical system over the wavelength range from 334nm to 650 nm was designed, providing good imaging performance in Modulation Transfer Function (MTF) 0.77 at 22.52/mm Nyquist frequency of the ICCD detector and 2.7μm in RMS spot radius.

| Type | Ritchey-Chrétien |
|---|---|
| Focal ratio $f$ | 11.4 |
| Effective $fl$ | 1140 mm |
| FOV | 17×17 arcmin² |
| Pixel resolution | 4 arcsec |
| Primary RC | 315 mm |
| Primary SC | -1.01 |
| Secondary RC | -63.8 mm |
| Secondary SC | -1.83 |

Table 2. Optical telescope characteristics

We use Zerodur® class 0 from SCHOTT for SMT mirrors, these being manufactured in collaboration with Samsung Electronics. Using a MRF machine, they were manufactured to within 0.017 waves (M1) and 0.020 waves (M2) in RMS WFE and the resulting surface quality meets the mirror tolerance budget

● Opto-Mechanics Design

The opto-mechanical concept for the telescope was two optical sub-assemblies (M1 and M2 sub-assemblies) to be fixed at the UFFO base plate via the M1 support plate mounted to the triangular mechanical brackets. M1 is mounted to the M1 support plate with 3 bi-pod type flexures. The M2 cell is mounted to the M1 support plate through 4 spider arms for reduction of an unnecessary stress from UFFO mechanics. The telescope was designed for the minimum obscuration ratio down to 12.5% (Figure 4)

A Finite Element Analysis (FEA) model of telescope was also constructed. Triangle type curvature based mesh was used and the constraints were given to the bottom plate joined with bolts. The results of static gravity analysis are shown in Table 3. Launch and operation directions are in parallel with Z direction of load. The analysis proves that it is stiff enough to survive from a launch stress about 5 times larger than the von Mises stress.

| Direction of Load | X | Y | Z |
|---|---|---|---|
| Minimum Safety Factor | 3.06 | 3.66 | 5.66 |

Table 3. Static analysis results with 60g loads

The frequency analysis was also performed. The first 5 resonance frequencies are listed in Table 4 and the first resonance arises from around 135Hz. It is clear that the first frequency is well above the typical frequency requirement, 50Hz [8], ensuring that our instrument will survive during the launch.

| Mode | NF (Hz) | Cycle (sec) |
|---|---|---|
| 1 | 135.12 | 0.007401 |
| 2 | 242 | 0.004132 |
| 3 | 251.05 | 0.003983 |
| 4 | 262.02 | 0.003817 |
| 5 | 327.44 | 0.003054 |

Table 4. Resonance frequency of SMT telescope

UFFO-P will be on a sun-synchronous orbit at 550km in altitude with 89° in inclination angle and will experience 15 thermal cycles / day. The temperature of UFFO-P will be controlled to be -10°C ~ 40°C at the bottom plate. If the satellite is properly operated, UFFO-P will be exposed to 0±7°C. It is of paramount importance to keep the distance between the primary and secondary mirrors within 5μm at any given temperature within the range. For this, we used invar for the 4 spider arms. To minimize the temperature effects on the primary mirror caused by a coefficient of thermal expansion (CTE) mismatch between the invar bipod flexure and the aluminum back structure, the bipod legs are designed to be 12mm long, 1.5mm wide, and 1.5mm thick and all secondary mirror support structure parts are made of invar. They are mounted to an invar cell with three flexure blades of 11mm long, 3mm wide, and 0.51mm thick. The results of thermal static analysis are listed in Table 5 and are clear that it satisfies the tolerance.

| Parameter | Unit | Tolerance | Analysis result (Abs.) |
|---|---|---|---|
| X decenter | mm | ±2.00E-02 | 2.58E-04 |
| Y decenter | mm | ±2.00E-02 | 7.26E-03 |
| X tilt | degrees | ±5.00E-02 | 1.19E-03 |
| Y tilt | degrees | ±5.00E-02 | 5.70E-04 |
| Z defocus | mm | ±5.00E-03 | 4.21E-06 |
| RMS WFE | waves | 1/20 | 1/40 |

Table 5. Thermal static analysis result for Δ30°C

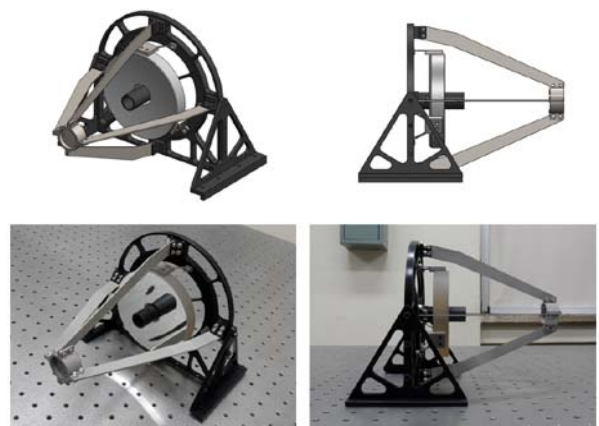

Figure 4. 3D model of SMT telescope (above) and Manufactured view (below)

UFFO-P is to have four parts for a baffle system i.e. a shade, blocking the incident light from outside the FOV; tube baffle, primary/secondary mirror baffle and detector baffle. The baffle structure design is in progress as of today.

- **Assembly and Performance Measurements**

Mirrors were bonded to the bi-pod type flexures using EC2216 B/A epoxy. It was cured at room temperature over 7 days. The SMT optical alignment was carried out using a WYCO RTI6100 interferometer. The integration setup for alignment is shown in Figure 5.

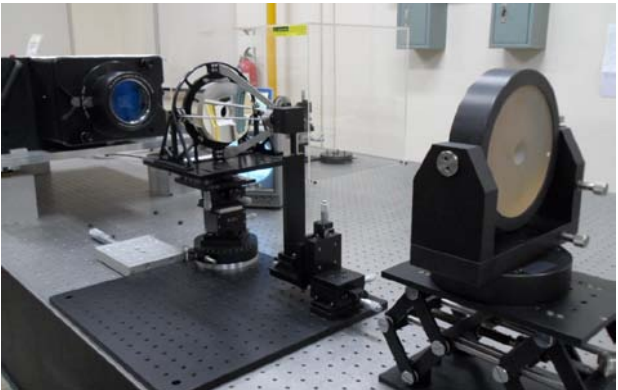

Figure 5. Experiment setup for alignment and Measurement of WFE

After completion of the alignment, the system WFE was measured to RMS 0.05 waves for 632.8 nm of wavelength (Figure 6). The final performance implies that the Integration and Test process was well controlled within the tolerances. The results satisfy the optical requirements of UFFO-P.

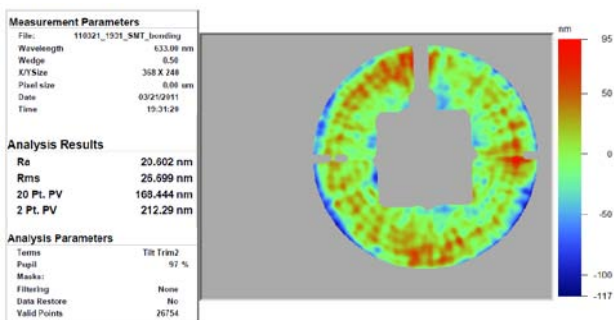

Figure 6. Measured system WFE of SMT telescope

- **Static Load Test**

The static load test was performed to evaluate survivability of the SMT structures during the launch. 2kg$f$ loading was applied to the secondary mirror cell structure of 40 g in mass, simulating the launch stress of 50 g. The resulting displacement curve is shown in Figure 7, exhibiting typical, but satisfactory hysteresis characteristics.

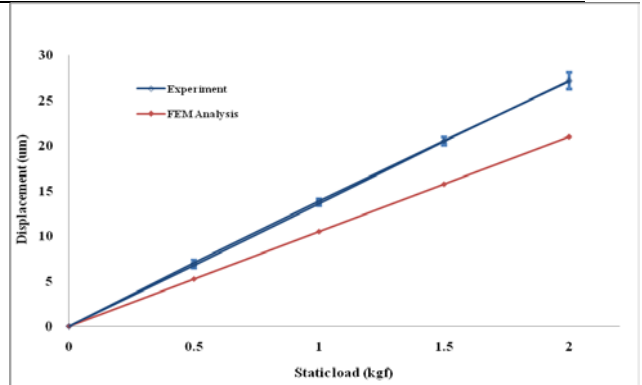

Figure 7. Structural displacement of SMT against static load

The resulting displacement increases almost linearly with the applied load as predicted from the FEA analysis. Whilst such displacement behavior is about 23% larger than the predicted one from FEA, the structure shows repeatable up and down hysteresis curves and their differences are within around 0.24μm. This is well below the tolerance budget the for secondary mirror displacement (20μm). From frequency and static loading analysis together with the experimental displacement measurement, we are now relatively confident that the telescope structure would survive well under the launch stresses.

## 3 Summary

The SMT optics system has a motorized slewing mirror subassembly to cover 35 degrees in the sky and its telescope has 17×17 arcmin$^2$ FOV with < 0.25 waves in RMS wave front error. All parts such as primary mirror, secondary mirror, opto-mechanical parts were designed, fabricated, and integrated to meet the requirements successfully. The integrated SMT system has RMS 0.05 wave for 632.8 nm in wavelength. According to gravity static analysis, resonance frequency analysis, thermal static analysis and static load test result, we are confident that the SMT optics system will survive well during the launch and in space operation. In July 2011, space environment tests such as shock, vibration and thermal cycling are to be carried out in National SPace Organization (NSPO), Taiwan and the results will be reported elsewhere.